\documentclass[letterpaper,twocolumn,10pt]{article}
\usepackage{usenix,epsfig,endnotes}
\usepackage{todonotes,graphicx,tabu,tabularx}
\usepackage{xspace}
\usepackage{hyperref}
\usepackage{booktabs}
\usepackage{cite}
\usepackage{balance}

\usepackage[inline]{enumitem}

\graphicspath{ {images/} }
\usepackage{color}
\begin{document}

\newcommand{\CloudCoaster}{\emph{CloudCoaster\xspace}\xspace}
\newcommand{\sysname}{\emph{CloudCoaster}\xspace}
\newcommand{\tian}[1]{{\color{red}\textbf{Tian: \textit{#1}}}}
\newcommand{\sam}[1]{{\color{blue}\textbf{Sam: \textit{#1}}}}

\date{}

\title{\Large \bf \CloudCoaster: Transient-aware Bursty Datacenter Workload Scheduling}

\author{
{\rm Samuel S. Ogden \qquad Tian Guo}\\
Worcester Polytechnic Institute \\
\{ssogden, tian\}@wpi.edu
} 

\maketitle

\thispagestyle{empty}

\subsection*{Abstract}

	Today's clusters often have to divide resources among a diverse set of jobs.  These jobs are heterogeneous both in execution time and in their rate of arrival.  Execution time heterogeneity has lead to the development of hybrid schedulers that can schedule both short and long jobs to ensure good task placement.  However, arrival rate heterogeneity, or \textit{burstiness}, remains a problem in existing schedulers.  These hybrid schedulers manage resources on statically provisioned cluster, which can quickly be overwhelmed by bursts in the number of arriving jobs.\par
	In this paper we propose \sysname, a hybrid scheduler that dynamically resizes the cluster by leveraging cheap transient servers.  \sysname schedules jobs in an intelligent way that increases job performance while reducing overall resource cost.  We evaluate the effectiveness of \sysname through simulations on real-world traces and compare it against a state-of-art hybrid scheduler.  \sysname improves the average queueing delay time of short jobs by 4.8X while maintaining long job performance. In addition, \sysname reduces the short partition budget by over 29.5\%.

\section{Introduction}

	Modern cluster workloads have to share resources among many diverse jobs.  These jobs can vary greatly in run duration, latency constraints and the number of tasks they consist of.  These variations prove challenging for the job schedulers that need to determine how to allocate resources efficiently.

	Previous work~\cite{eagle.Delgado:2016:JSE:2987550.2987563,Hawk.191591,google_traces.Reiss:2012:HDC:2391229.2391236} shows that the run duration of jobs is heterogeneous in modern workloads, with the average duration of short and long jobs being orders of magnitude different.  Adding to this, many of the short jobs are user interactive and thus have tight latency requirements.  
		To address these issues while minimizing scheduling delay, hybrid schedulers such as Eagle~\cite{eagle.Delgado:2016:JSE:2987550.2987563} and Mercury~\cite{Mercury.190440} divide cluster resources and schedule tasks by taking into account of job heterogeneity. 

However, modern cluster workloads exhibit tendency of resource needs increasing and decreasing in bursts, caused by dynamic job arrival rates and job sizes. As an example, a study~\cite{google_traces.Reiss:2012:HDC:2391229.2391236} on Google cluster trace shows that even though on average jobs only have 35 tasks, some may have as many as 50 thousand tasks. Furthermore, jobs themselves do not arrive at a constant rate, leading to phases of over- and under-utilization of cluster resources. 

To deal with dynamic workload cost-effectively, one approach is to adjust the cluster size accordingly~\cite{2011socc:clustersize}. The key questions for using dynamic clusters are when to add and remove servers, as well as which servers to use. Besides traditional on-demand servers, cloud providers~\cite{ws:amazon_spot,ws:google_preemptible} started to offer a new server type called transient servers. These transient servers provide significant cost saving~\cite{183100,Chohan:2010:SSR:1863103.1863110,flint.Sharma:2016:FBD:2901318.2901319}, compared to using on-demand resources, but are subject to unpredictable availability upon revocations. In this work, we ask the question: how we could leverage cheap transient servers to handle bursty cluster workload.

Towards answering this question, we introduce \sysname, a new hybrid scheduler that adapts cluster resources through efficient usage of transient servers in order to cope with variations in task arrival rates.  In designing this system we make the following contributions.  First, we introduce a new cluster utilization metric, the \textit{long load ratio}.  Next, we propose a transient aware resource management algorithm that utilizes this metric to intelligently add and remove resources.  Finally, we show through simulation the feasibility of our system in decreasing short job queueing delay by up to 4.8X.
	
\section{Background and Motivation}

One of the key challenges in designing job schedulers is to account for the heterogeneous and bursty nature of modern workloads. State-of-the-art schedulers often assume statically provisioned clusters and therefore might not fully benefit from additional transient servers. In this section we briefly discuss the underlying challenges to designing a transient-aware hybrid scheduler for dynamic-sized cloud clusters.  

\subsection{Job Heterogeneity}
	It is common for a cluster to process workloads in which jobs have large variations in run duration, resource need and latency requirements.  Prior work~\cite{Hawk.191591} has found that in some extreme cases just 2\% of jobs can account for over 99\% of cluster time.  Due to their long run time, these jobs are termed long jobs, while the remaining jobs are termed short jobs.
	
    	Furthermore, short jobs are typically associated with user interaction and thus are highly latency sensitive.  On the other hand, long jobs tend to be batch jobs and therefore less concerned with latency but require good placement for optimum runtime.  Based on this difference, short jobs are generally best scheduled \emph{quickly} by decentralized schedulers like Sparrow~\cite{Sparrow.Ousterhout:2013:SDL:2517349.2522716}, while long jobs are generally best served by centralized schedulers such as YARN~\cite{YARN.Vavilapalli:2013:AHY:2523616.2523633}.

\subsection{Hybrid Schedulers}
	Due to the specialized niches of centralized and decentralized schedulers a new class of schedulers, called hybrid schedulers, has been developed.  These schedulers generally use a combination of a single centralized scheduler for long tasks,  and a number of decentralized schedulers to quickly place short tasks.  This allows hybrid schedulers to achieve low scheduling delays for short task and high quality placements for long tasks.
	
	One hybrid scheduler that is of particular interest is Eagle~\cite{eagle.Delgado:2016:JSE:2987550.2987563}.  Eagle uses decentralized schedulers to gather hints about servers that have long tasks, and leverages this information to avoid scheduling short tasks behind long ones. In addition, Eagle creates a small cluster partition \emph{only} for short jobs---this ensures that no short tasks will be scheduled behind long tasks, even during times of high cluster utilization. However, short tasks might still experience long queueing delays when a large number of long tasks occupy the cluster and all incoming short jobs are placed into the short-job only partition. 

\subsection{Burstiness of Workload}
	The potential for high cluster utilization is caused by non-uniformity in the rate of task arrival, which has not been previously addressed by hybrid schedulers.  Jobs can arrive at irregular intervals and each job can have a widely varying number of tasks.  For example, within the publicly available Google cluster trace~\cite{google_traces.Reiss:2012:HDC:2391229.2391236} the number of tasks per job varies from one to 49960.
	
\begin{figure}[t]
	\centering
    	\includegraphics[width=\columnwidth ]{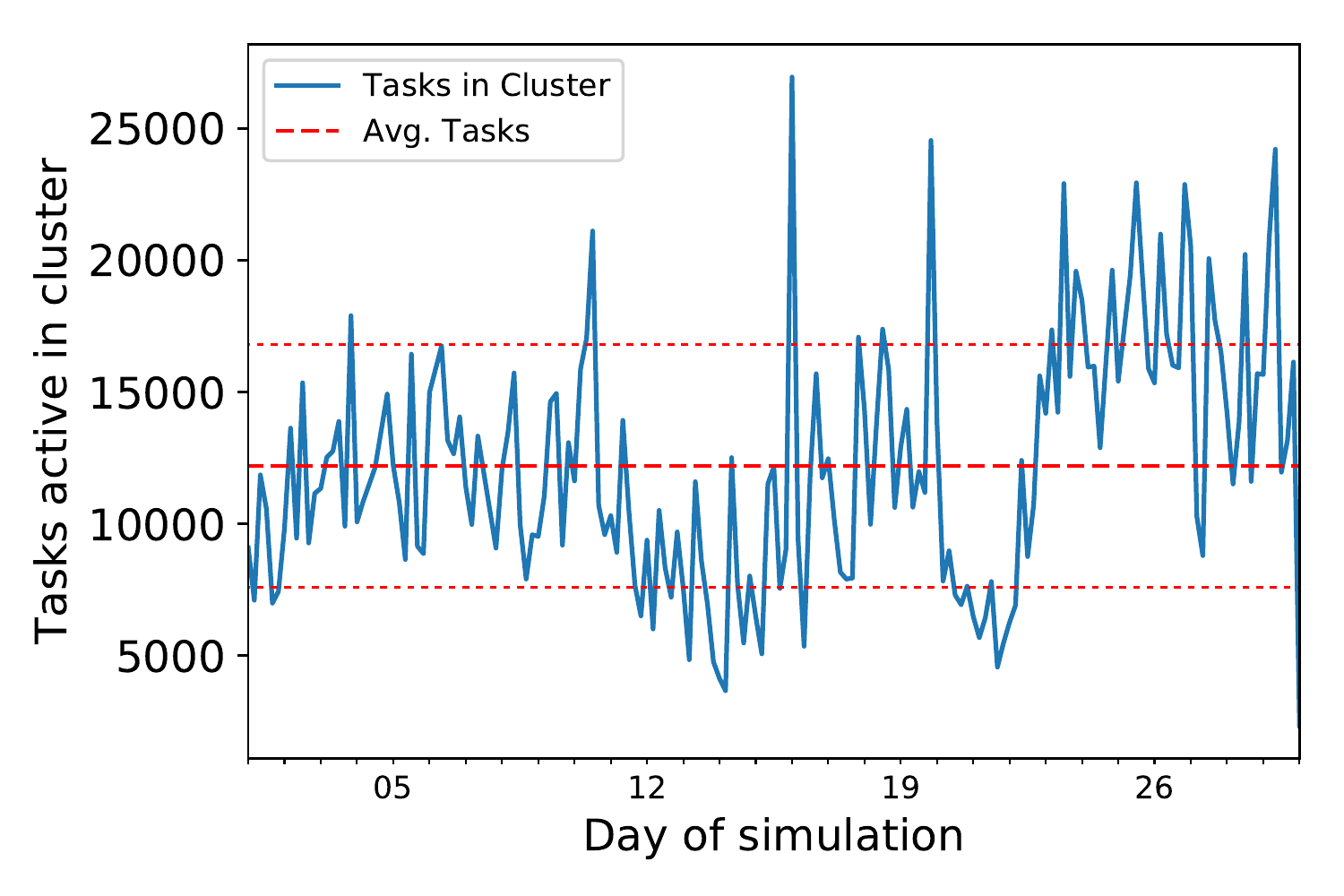}
	\caption{\textbf{Theoretical number of concurrent tasks from the Google trace~\cite{google_traces.Reiss:2012:HDC:2391229.2391236}.} \emph{Here, we assume a cluster with unlimited resources and an omniscient scheduler with no scheduling delay. The number of tasks is averaged first over 100-seconds and then over 4-hour periods to improve readability.  The average and standard deviation are represented in red dashed lines.  Large spikes and troughs can be seen, indicating workload variations.}}
       	\label{GoogleConcurentJobs4H}
\end{figure}

	This large variation in the number of tasks entering the cluster leads to large variations in the amount of resources needed. We analyze the same Google cluster trace for the number of concurrent jobs that would be running in a cluster, with unlimited resources and an omniscient scheduler. Figure~\ref{GoogleConcurentJobs4H} shows that number of active tasks varies significantly, more than 6X differences between the highest and lowest points. This indicates that not only are there times when extra resources would be beneficial but also times when the cluster is effectively under-utilized.
	
\subsection{Transient Servers}\label{Background.TransientServers}

	Many cloud providers such as Amazon Web Services~\cite{AWS} and Google Compute Engine~\cite{GCE} now offer transient servers.  These are servers that are leased at a relatively lower unit price but with no availability guarantees. In the Amazon's case that employs dynamic pricing, cloud customers bid for transient servers by specifying how much they are willing to pay for resources.  If this bid is above the current transient market price, then they are granted these transient servers.  These transient servers are available to the bidder until the market price rises above her bid.  At this point the customer is given a warning and after a short period these resources are revoked by the cloud provider. 
	
Previous work has shown that transient servers can often a discount of up to 90\%~\cite{ws:amazon_spot} and with an effective average cost of only 30\% of their on-demand counterparts~\cite{flint.Sharma:2016:FBD:2901318.2901319}. Transient servers' low cost makes them attractive for supplementing cluster's dynamic resource demands, even with their inherent unpredictable availability.

\section{System Design}

	\emph{CloudCoaster}\endnote{Simulation code for \sysname is available on GitHub at {\tt https://github.com/samogden/CloudCoaster}} is a hybrid scheduler that works with dynamically-sized clusters of both on-demand and transient servers. More specifically, we use transient servers to grow and shrink the cluster to more closely match the supply of resources to workload demand, so as to improve short task response time.  We focus on the short tasks rather than long tasks for two reasons.  First, short tasks are more sensitive to the resource usage of other tasks running in the cluster. For example, if the current cluster is occupied with long tasks, short tasks might experience head-of-line blocks~\cite{eagle.Delgado:2016:JSE:2987550.2987563} if being scheduled behind these long tasks. Second, short tasks are in a better position to exploit transient resources as they are more likely to complete before the revocation of transient servers. 
	
\begin{figure}[t]
	\centering
    	\includegraphics[width=\columnwidth ]{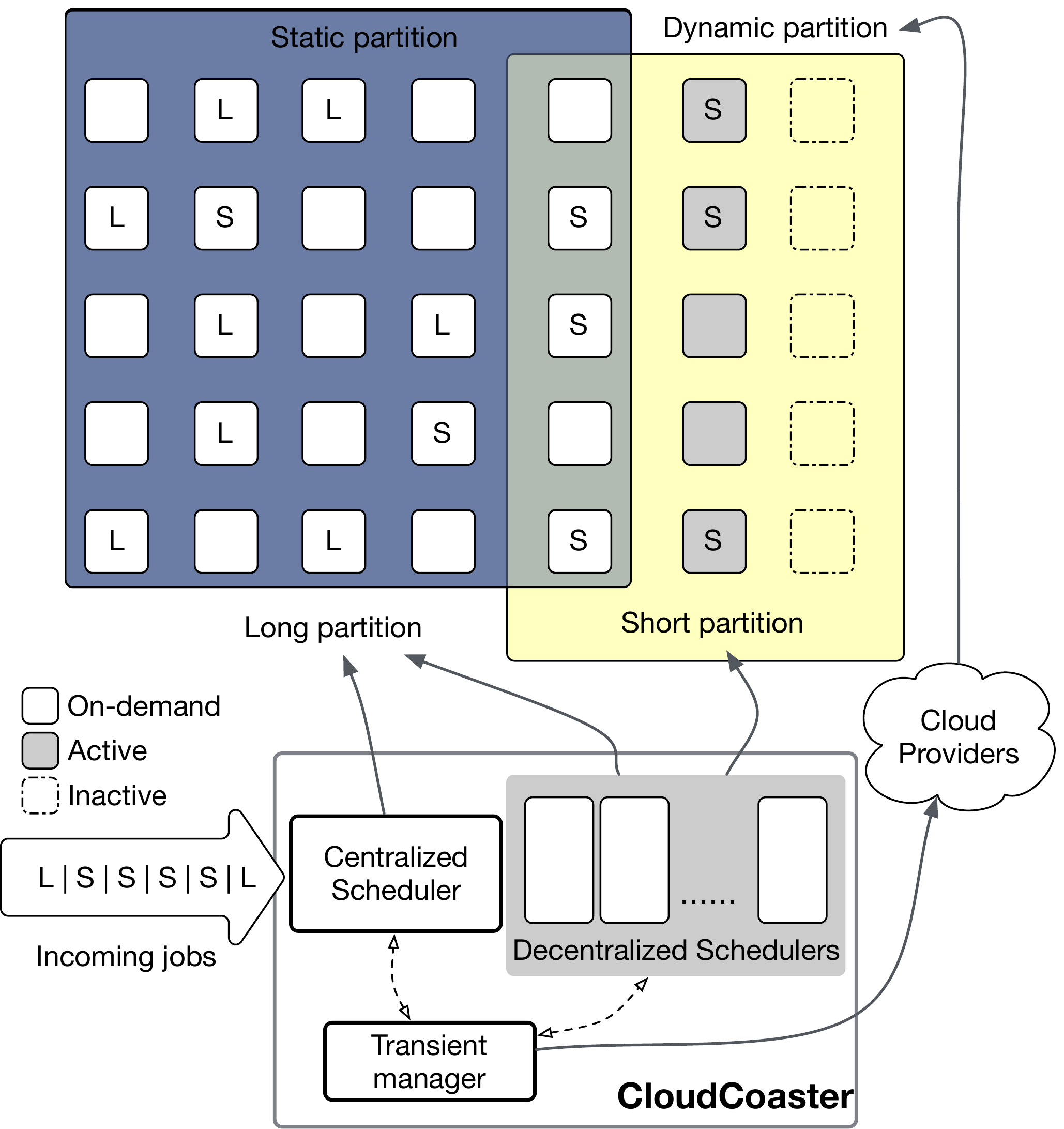}
	\caption{\textbf{\sysname overview.}  \emph{Incoming jobs arrive through the centralized scheduler which passes short jobs off to decentralized schedulers.  The transient manager monitors the state of the cluster, requesting and releasing resources from the cloud provider as needed.}}
       	\label{CloudCoasterOverview}
\end{figure}

	\sysname is based on a state-of-the-art hybrid scheduler~\cite{eagle.Delgado:2016:JSE:2987550.2987563}, and utilizes the same centralized/decentralized paradigm. We depict our overall design of \sysname in Figure~\ref{CloudCoasterOverview}. \sysname also partitions the cluster, in which long partition can be used for both long and short jobs, while short partition is only used for short jobs. \sysname dynamically resize the short partition using cheap transient servers. This is achieved by the \textit{Transient Manager}  that obtains information from the centralized scheduler in order to derive the cluster state.  From this state, \sysname makes informed decisions about whether to allocate or deallocate resources.  The goal of this dynamic resizing is to minimize the effects of burstiness on \emph{short job} queueing time.  By leveraging cheap transient resources, \sysname is able to increase the number of servers available to short jobs without exceeding the cost budget.  Additionally, using this cluster state, \sysname determines when the cluster is less crowded and releases unneeded resources, thus reducing budget.  
	
In sum, our design of \sysname is influenced by the following three questions: 
	\begin{enumerate*}
	\setlength\itemsep{1em}
		\item[\textbf{(1)}]{How many transient servers can we supplement given a set operational budget?}
		\item[\textbf{(2)}]{When and how should be we resize the cluster with transient servers?}
		\item[\textbf{(3)}]{How do we efficiently utilize transient servers based on their unique cost and availability characteristics?}
	\end{enumerate*}

\subsection{On-demand Server Replacement} \label{DynamicClusterResizing}
	The key insight of \sysname is that statically provisioned clusters will cause both resource scarcity and surplus due to workload burstiness.  To mitigate the shortcomings of statically provisioned clusters cost-effectively, we propose using transient resources to increase capacity at times of high utilization and reduce capacity at times of low usage.

	Figure~\ref{CloudCoasterOverview} shows how \sysname uses a static partition in conjunction with a dynamic short partition for running both long and short jobs. The static partition is comprised of a fixed number of on-demand servers, including a small number that is dedicated to running short tasks. These designated on-demand servers act as a buffer to absorb the unpredictability of transient servers.  Further, the dynamic partition is reserved for running short jobs and can consist of a varying number of transient servers. Due to the cost benefits of transient servers, the dynamic partition may contain many more servers than the static partition for the same cost.
	
	More concretely, if we define the the individual cost of on-demand and transient servers as $c_{static}$ and $c_{trans}$ respectively, then we can define their cost ratio, 
	$$r=\frac{c_{static}}{c_{trans}}.$$
	This value $\lfloor r \rfloor$ is the number of transient servers that can be used for the cost of one equivalent on-demand server.  This value is generally in the range [1,10] with a reasonable value being $r=3$~\cite{flint.Sharma:2016:FBD:2901318.2901319}.  Let us denote the number of on-demand servers we have in the short-only partition of a purely static cluster as $N$, and the percentage of the on-demand servers that we could replace with transient servers as $p$. We can then calculate the maximum number of transient servers that are available to \sysname as $ K = r N p $. As such, \sysname can manage a short-only partition of size $T$ where: 
	
	$$T =  r N p + (1 - p)N  = N ( (r-1)p + 1).$$
	
	For example, if we were to convert 50\% of our on-demand servers from the short-only partition to transient servers, we could end up with a total of $T = N ( (3-1)(0.5) + 1) = 2N$ servers in our short-only partition.

\subsection{Short Partition Resizing}
	The full potential of transient servers is only realized when they are added and removed as needed.  To do this \sysname tracks the state of the cluster through a metric called the \emph{long-load ratio}. This ratio  $l_r$ is calculated as the number of servers processing long tasks divided by the total number of servers in the cluster.  That is, 
	$$l_r = \frac{N_{long}}{N_{total}},$$

where $N_{long}$ is the number of of servers with long tasks and $N_{total}$ is the total number of servers.  This metric indicates the likelihood that a randomly assigned short task would be enqueued behind a long task, an undesirable situation.  Based on this, higher values of $l_r$ mean that we should increase the size of the short partition, thereby increasing the number of servers available to short tasks.
	
	When \sysname first comes online, $l_r$ is set to zero. Whenever a long job enters, \sysname recalculates $l_r$ based on the long job's tasks assignments. Similarly, when a long tasks exits the cluster, $l_r$ is also updated. After every recalculation of $l_r$, \sysname compares $l_r$ with a predefined threshold $L_r^T$. If $l_r$ is above the threshold, \sysname adds a transient server. Otherwise, \sysname removes one transient server. When releasing the transient server, \sysname instructs the server to complete all of its currently enqueued tasks before shutting down. 
	
	The process of adding or removing transient servers is repeated until either $l_r = L_r^T$ or we can not further request or release transient servers due to constraints such as cost budget. In sum, \sysname only updates $l_r$ whenever a long task enters or exists the cluster or a transient server is added or removed. 
	
\subsection{Discussions}

When designing \sysname, we take the approach to more aggressively increasing and more conservatively decreasing the number transient servers. Such design is driven by two aspects: first, quickly increasing the number in the cases of many long jobs ensures that short jobs will have access to enough servers. Second, the slow removal of resources allows avoiding the non-negligible provisioning time in spite of fluctuating workloads.

When using transient servers in dynamic clusters, one needs to be aware of two complications. First, as a cheaper alternative to on-demand servers, transient servers can be revoked by cloud providers at any time with a short time window, e.g., 30 seconds~\cite{}. As such, transient-aware schedulers might need to reschedule short tasks that were only enqueued on transient servers---leading to unexpected delays or even missing tasks. To avoid rescheduling, in this work, \sysname ensures that at least one copy of the short tasks is scheduled to an on-demand server. Second, some types of transient servers might not be available upon being requested~\cite{2016icdcs:spotlight}. To overcome this availability issue, \sysname follows similar strategies~\cite{SpotCheck.Sharma:2015:SDD:2741948.2741953} to work with a wide variety of transient servers of different capacities and \emph{slice} them accordingly.  	

\section{Evaluation}

	To evaluate \sysname, we ran simulations using the Yahoo trace~\cite{yahoo_traces.Chen:2011:CEM:2060110.2060856, Hawk.191591}. This allows us to investigate three key aspects: (1) the effectiveness of \sysname in adjusting short-only partition dynamically; (2) the impact of transient server lifetimes on execution of short jobs; (3) and the cost reductions from using transient servers.
	
	Unless otherwise noted, our simulations use a baseline of a cluster with 4000 on-demand servers, 80 of which are used for short jobs. That is, we have $N_s = 80$ and the budget to dynamically adjust the size of the short-only partition is constrained by the total cost of $N_s$ servers. Further, we configure $p=0.5$ to allow up to half of short-only partition be replaced with transient servers. We vary the cost ratio between on-demand and transient servers from one to three. As such, \sysname can use up to 40, 80 and 120 transient servers, respectively.The replacement threshold was $L_r^T=0.95$ and provisioning delay of transient servers was set to 120 seconds.

\subsection{Effectiveness of Dynamic Partition}

\begin{figure}[t]
	\centering
    	\includegraphics[width=\columnwidth ]{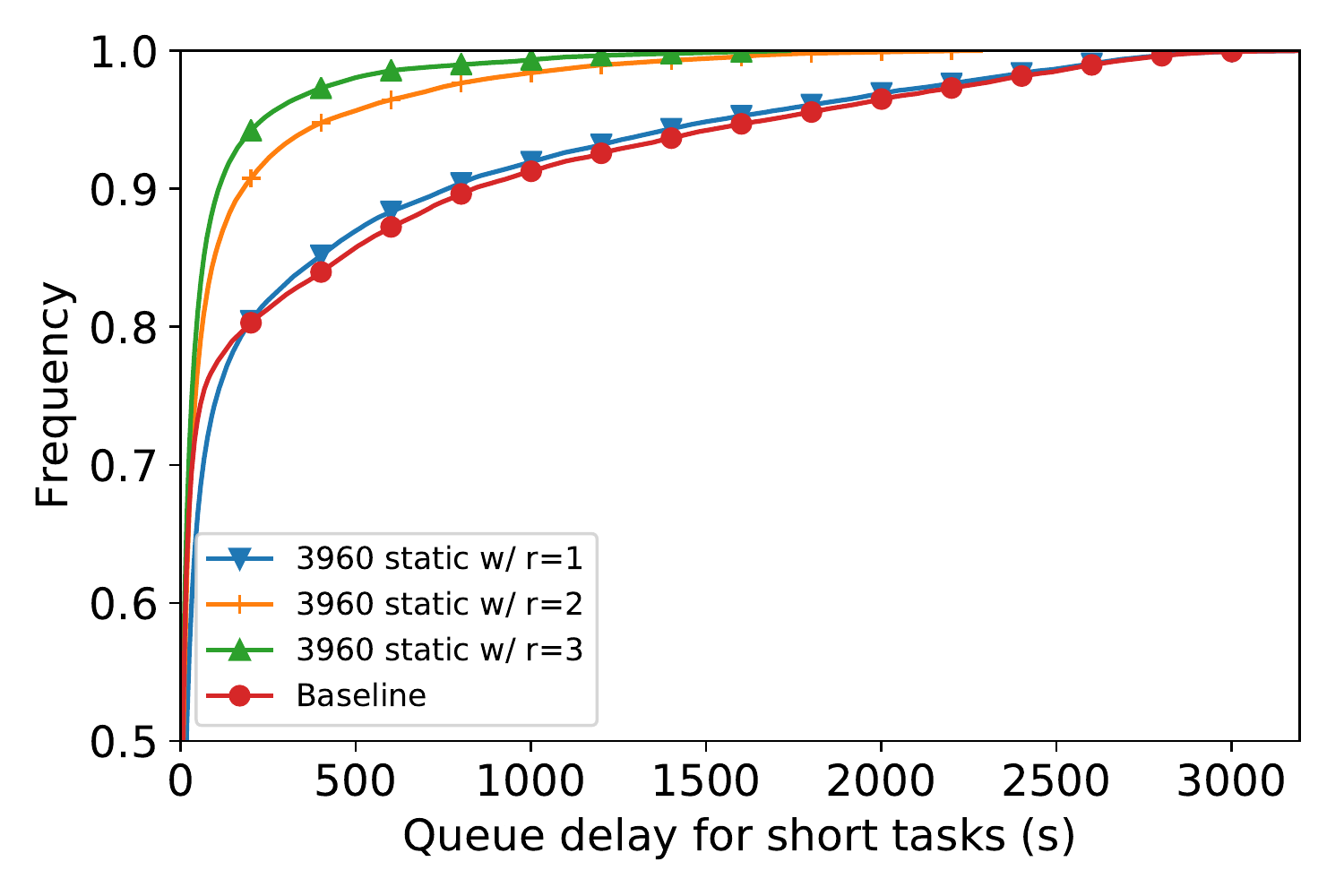}
	\caption{\textbf{CDFs of short tasks queueing delay.} \emph{We replaced on-demand servers with transient servers in Yahoo traces\cite{Hawk.191591}.  The baseline had 80 on-demand servers, 40 of which were replaced.  Replacements were done with $r=1,2,3$ and a threshold $L_r^T=0.95$.}}
       	\label{YahooReplacementCDFs}
\end{figure}

	We first investigate the effectiveness of using cheaper transient servers whenever needed, under different cost ratios. Figure~\ref{YahooReplacementCDFs} compares the CDFs of the queueing delay of short tasks. As we can see, using transient servers when they cost the same as on-demand counterparts ($r=1$) leads to similar performance of using a state-of-the-art hybrid scheduler Eagle~\cite{eagle.Delgado:2016:JSE:2987550.2987563}, labeled as \emph{Baseline}. The slight difference is mainly caused by removing transient servers and the provisioning overhead associated with adding transient server. 
	
	Second, we observe that as transient servers become cheaper ($r=2$ and $r=3$), \sysname achieves significant improvement in queueing time. Concretely, the average (and maximum) queueing time is reduced from 232.3 (and 3194) seconds when using Eagle to 48.25 (and 1737) seconds---a 4.8X (and 1.83X) improvement---when the cost ratio is $r=3$. This demonstrates that \sysname is able to add in and remove transient servers appropriately, preventing short tasks to be adversely affected by a large number of long tasks present in the cluster. Further, as the cost ratio increases, we expect to see the queueing delay of short tasks improve further.

	\textbf{Results:} \emph{\sysname's use of long-load ratio to dynamically adjust short-only partition is effectiveness. Replacing on-demand resources with transient instances improves average short job queueing delay by up to 4.8X.}

\subsection{Transient Server Lifetimes and Cost Analysis}

Next, to understand the potential impact of transient server revocations, we examine the amount of time each transient servers were used in our simulation. Table~\ref{TransientServerTable} shows that the average lifetimes of transient servers is between 0.77 to 0.82 hours. This suggests a low probability of involuntary revocations given that the mean-time-to-failure of transient servers offered by Amazon and Google can be significantly more than 18 hours~\cite{flint.Sharma:2016:FBD:2901318.2901319}. Moreover, the longest time a transient server was used in our simulation is 12.8 hours---still much smaller than reported revocation frequency. 

Next, we quantify the cost-effectiveness of \sysname's dynamic short-only partition. To do so, \sysname keeps track of the number of active transient servers periodically. In Table~\ref{TransientServerTable}, we show the average number of transient servers used throughout our simulations. Then we calculate the average number of \emph{r-normalized} on-demand servers by dividing the corresponding $r$ value. Compared to the baseline of 40 on-demand servers when using Eagle as the scheduler, \sysname uses up to 11.8 less on-demand servers on average. This translates to a cost saving of 29.5\% in operating short-only partition. 

\textbf{Results:} \emph{In our simulations, \sysname only uses transient servers for much shorter time than the real-world mean-time-to-failure of transient servers. In addition, \sysname achieves up to 29.5\% cost savings for short-only partition.} 
 
\begin{table}[]
\centering
\resizebox{\columnwidth}{!}{%
\begin{tabular}{@{}r|r|r|r|r@{}}
\cmidrule(l){2-5}
 & \multicolumn{2}{r|}{\textbf{Active time (hours)}} & \multicolumn{2}{r|}{\textbf{Active number}} \\ \midrule
$r$ & Average & Maximum & \begin{tabular}[c]{@{}r@{}}Average\\ transient\end{tabular} & \begin{tabular}[c]{@{}r@{}}$r$-normalized \\ avg. on-demand\end{tabular} \\ \midrule
1 & 0.77 & 12.8 & 29.0 & 29.0 \\
2 & 0.82 & 12.5 & 56.5 & 28.3 \\
3 & 0.79 & 12.5 & 84.5 & 28.2 \\ \bottomrule
\end{tabular}%
}
\caption{\textbf{Number of transient servers and their lifetimes in our simulation.} \emph{We analyze the average and maximum amount of time transient servers were used in our simulation. In all cases, \sysname uses transient servers for less than the historical mean-time-to-failure of Amazon's spot instances. In addition, \sysname uses up to 29 on-demand servers, compared to 40 in the baseline.}}
\label{TransientServerTable}
\end{table}

\section{Related Works}

\textbf{Schedulers.}
	Schedulers have long been an active field of research as they are a critical part of any clusters.  Centralized schedulers such as YARN~\cite{YARN.Vavilapalli:2013:AHY:2523616.2523633} and others~\cite{Mesos.Hindman:2011:MPF:1972457.1972488,Isard:2009:QFS:1629575.1629601,Ferguson:2012:JGJ:2168836.2168847,Zaharia:2010:DSS:1755913.1755940,199390} assign each job optimally but slowly and are overwhelmed by the sheer volume of short jobs in modern applications. Decentralized schedulers such as Sparrow~\cite{Sparrow.Ousterhout:2013:SDL:2517349.2522716} excel at numerous short jobs but experience poor performance when faced with a heterogeneous workload.  Hybrid schedulers~\cite{Mercury.190440, Hawk.191591, eagle.Delgado:2016:JSE:2987550.2987563, Boutin:2014:ASC:2685048.2685071} often divide jobs into different categories allowing more efficient scheduling. Our work builds atop a state-of-the-art hybrid scheduler to be transient-aware.  
	
\textbf{Transient servers.} The cheaper transient servers have attracted a lot of research attentions. Prior work~\cite{2016journal:optispot} that exploits the cost benefits ranges from running batch jobs~\cite{183100,2015socc:spoton} to interactive workloads~\cite{SpotCheck.Sharma:2015:SDD:2741948.2741953,flint.Sharma:2016:FBD:2901318.2901319}. The resource intensive nature of big data processing~\cite{Binnig2015SpotgresP,Salama:2015:CFP:2723372.2749437,2017eurosys:pado} and machine learning workloads~\cite{183100,Yan:2016:TTC:2987550.2987576,2019icac:measurement,flint.Sharma:2016:FBD:2901318.2901319,2017eurosys:proteus} also make them good candidates for running on transient servers. To mitigate the revocation impacts of transient servers, fault-tolerant techniques such as checkpointing~\cite{flint.Sharma:2016:FBD:2901318.2901319,2010cloud:spotcheckpoint,Yan:2016:TTC:2987550.2987576} and server selection policies~\cite{SpotCheck.Sharma:2015:SDD:2741948.2741953,2017sigmetrics:exosphere} have been explored extensively. 
	
\section{Conclusion and Future Work}

	In this paper we introduce \sysname, a hybrid scheduler that resizes the cluster to increase performance and decrease cost for short jobs.  The availability and low price of transient instances allow for extra resources to be added during times of high utilization and idle resources to be removed during times of low utilization.  This improves the average queueing delay of short tasks by 4.8X, while reducing costs of short-only partition by up to 29.5\%.
	
	As part of the future work we first plan to evaluate \sysname using large scale Google cluster traces.  Additionally we plan to implement \sysname as a scheduling plugin for Spark~\cite{Zaharia:2010:SCC:1863103.1863113} to further explore the cost benefits of transient servers.
	
\textbf{Acknowledgements.} This work is supported in part by
National Science Foundation grants \#1755659 and \#1815619,
and Google Cloud Platform Research credits.

\balance
{\footnotesize \bibliographystyle{acm}
\bibliography{bib.bib}}

\begin{thebibliography}{10}

\bibitem{AWS}
Amazon web services {(AWS)}.
\newblock \url{https://aws.amazon.com/}.
\newblock Accessed: 2018-02-06.

\bibitem{GCE}
Compute engine -- {IaaS}.
\newblock \url{https://cloud.google.com/compute/}.
\newblock Accessed: 2018-02-06.

\bibitem{Binnig2015SpotgresP}
{\sc Binnig, C., Salama, A., Zamanian, E., El-Hindi, M., Feil, S., and Ziegler,
  T.}
\newblock Spotgres - parallel data analytics on spot instances.
\newblock {\em 2015 31st IEEE International Conference on Data Engineering
  Workshops\/} (2015), 14--21.

\bibitem{Boutin:2014:ASC:2685048.2685071}
{\sc Boutin, E., Ekanayake, J., Lin, W., Shi, B., Zhou, J., Qian, Z., Wu, M.,
  and Zhou, L.}
\newblock Apollo: Scalable and coordinated scheduling for cloud-scale
  computing.
\newblock In {\em Proceedings of the 11th USENIX Conference on Operating
  Systems Design and Implementation\/} (Berkeley, CA, USA, 2014), OSDI'14,
  USENIX Association, pp.~285--300.

\bibitem{yahoo_traces.Chen:2011:CEM:2060110.2060856}
{\sc Chen, Y., Ganapathi, A., Griffith, R., and Katz, R.}
\newblock The case for evaluating mapreduce performance using workload suites.
\newblock In {\em Proceedings of the 2011 IEEE 19th Annual International
  Symposium on Modelling, Analysis, and Simulation of Computer and
  Telecommunication Systems\/} (Washington, DC, USA, 2011), MASCOTS '11, IEEE
  Computer Society, pp.~390--399.

\bibitem{Chohan:2010:SSR:1863103.1863110}
{\sc Chohan, N., Castillo, C., Spreitzer, M., Steinder, M., Tantawi, A., and
  Krintz, C.}
\newblock See spot run: Using spot instances for mapreduce workflows.
\newblock In {\em Proceedings of the 2Nd USENIX Conference on Hot Topics in
  Cloud Computing\/} (Berkeley, CA, USA, 2010), HotCloud'10, USENIX
  Association, pp.~7--7.

\bibitem{ws:google_preemptible}
{\sc Cloud, G.}
\newblock Preemptible virtual machines.
\newblock \url{https://cloud.google.com/preemptible-vms/}, Accessed on June
  2019.

\bibitem{eagle.Delgado:2016:JSE:2987550.2987563}
{\sc Delgado, P., Didona, D., Dinu, F., and Zwaenepoel, W.}
\newblock Job-aware scheduling in eagle: Divide and stick to your probes.
\newblock In {\em Proceedings of the Seventh ACM Symposium on Cloud
  Computing\/} (New York, NY, USA, 2016), SoCC '16, ACM, pp.~497--509.

\bibitem{Hawk.191591}
{\sc Delgado, P., Dinu, F., Kermarrec, A.-M., and Zwaenepoel, W.}
\newblock Hawk: Hybrid datacenter scheduling.
\newblock In {\em 2015 {USENIX} Annual Technical Conference ({USENIX} {ATC}
  15)\/} (Santa Clara, CA, 2015), {USENIX} Association, pp.~499--510.

\bibitem{2016journal:optispot}
{\sc Dubois, D.~J., and Casale, G.}
\newblock Optispot: minimizing application deployment cost using spot cloud
  resources.
\newblock {\em Cluster Computing 19}, 2 (Jun 2016), 893--909.

\bibitem{ws:amazon_spot}
{\sc EC2, A.}
\newblock Amazon ec2 spot instances---run fault-tolerant workloads for up to
  90\% off.
\newblock \url{https://aws.amazon.com/ec2/spot/}, Accessed on June 2019.

\bibitem{Ferguson:2012:JGJ:2168836.2168847}
{\sc Ferguson, A.~D., Bodik, P., Kandula, S., Boutin, E., and Fonseca, R.}
\newblock Jockey: Guaranteed job latency in data parallel clusters.
\newblock In {\em Proceedings of the 7th ACM European Conference on Computer
  Systems\/} (New York, NY, USA, 2012), EuroSys '12, ACM, pp.~99--112.

\bibitem{199390}
{\sc Gog, I., Schwarzkopf, M., Gleave, A., Watson, R. N.~M., and Hand, S.}
\newblock Firmament: Fast, centralized cluster scheduling at scale.
\newblock In {\em 12th {USENIX} Symposium on Operating Systems Design and
  Implementation ({OSDI} 16)\/} (Savannah, GA, 2016), {USENIX} Association,
  pp.~99--115.

\bibitem{2017eurosys:proteus}
{\sc Harlap, A., Tumanov, A., Chung, A., Ganger, G.~R., and Gibbons, P.~B.}
\newblock Proteus: Agile ml elasticity through tiered reliability in dynamic
  resource markets.
\newblock In {\em Proceedings of the Twelfth European Conference on Computer
  Systems\/} (New York, NY, USA, 2017), EuroSys '17, ACM, pp.~589--604.

\bibitem{2011socc:clustersize}
{\sc Herodotou, H., Dong, F., and Babu, S.}
\newblock No one (cluster) size fits all: Automatic cluster sizing for
  data-intensive analytics.
\newblock In {\em Proceedings of the 2Nd ACM Symposium on Cloud Computing\/}
  (New York, NY, USA, 2011), SOCC '11, ACM, pp.~18:1--18:14.

\bibitem{Mesos.Hindman:2011:MPF:1972457.1972488}
{\sc Hindman, B., Konwinski, A., Zaharia, M., Ghodsi, A., Joseph, A.~D., Katz,
  R., Shenker, S., and Stoica, I.}
\newblock Mesos: A platform for fine-grained resource sharing in the data
  center.
\newblock In {\em Proceedings of the 8th USENIX Conference on Networked Systems
  Design and Implementation\/} (Berkeley, CA, USA, 2011), NSDI'11, USENIX
  Association, pp.~295--308.

\bibitem{Isard:2009:QFS:1629575.1629601}
{\sc Isard, M., Prabhakaran, V., Currey, J., Wieder, U., Talwar, K., and
  Goldberg, A.}
\newblock Quincy: Fair scheduling for distributed computing clusters.
\newblock In {\em Proceedings of the ACM SIGOPS 22Nd Symposium on Operating
  Systems Principles\/} (New York, NY, USA, 2009), SOSP '09, ACM, pp.~261--276.

\bibitem{Mercury.190440}
{\sc Karanasos, K., Rao, S., Curino, C., Douglas, C., Chaliparambil, K.,
  Fumarola, G.~M., Heddaya, S., Ramakrishnan, R., and Sakalanaga, S.}
\newblock Mercury: Hybrid centralized and distributed scheduling in large
  shared clusters.
\newblock In {\em 2015 {USENIX} Annual Technical Conference ({USENIX} {ATC}
  15)\/} (Santa Clara, CA, 2015), {USENIX} Association, pp.~485--497.

\bibitem{2019icac:measurement}
{\sc Li, S., Walls, R., Xu, L., and Guo, T.}
\newblock Speeding up deep learning with transient servers.
\newblock In {\em The16th IEEE International Conference on Autonomic Computing
  ({ICAC})\/} (Umeå, Sweden, 2019), IEEE.

\bibitem{183100}
{\sc Menache, I., Shamir, O., and Jain, N.}
\newblock On-demand, spot, or both: Dynamic resource allocation for executing
  batch jobs in the cloud.
\newblock In {\em 11th International Conference on Autonomic Computing ({ICAC}
  14)\/} (Philadelphia, PA, 2014), {USENIX} Association, pp.~177--187.

\bibitem{Sparrow.Ousterhout:2013:SDL:2517349.2522716}
{\sc Ousterhout, K., Wendell, P., Zaharia, M., and Stoica, I.}
\newblock Sparrow: Distributed, low latency scheduling.
\newblock In {\em Proceedings of the Twenty-Fourth ACM Symposium on Operating
  Systems Principles\/} (New York, NY, USA, 2013), SOSP '13, ACM, pp.~69--84.

\bibitem{2016icdcs:spotlight}
{\sc {Ouyang}, X., {Irwin}, D., and {Shenoy}, P.}
\newblock Spotlight: An information service for the cloud.
\newblock In {\em 2016 IEEE 36th International Conference on Distributed
  Computing Systems (ICDCS)\/} (June 2016), pp.~425--436.

\bibitem{google_traces.Reiss:2012:HDC:2391229.2391236}
{\sc Reiss, C., Tumanov, A., Ganger, G.~R., Katz, R.~H., and Kozuch, M.~A.}
\newblock Heterogeneity and dynamicity of clouds at scale: Google trace
  analysis.
\newblock In {\em Proceedings of the Third ACM Symposium on Cloud Computing\/}
  (New York, NY, USA, 2012), SoCC '12, ACM, pp.~7:1--7:13.

\bibitem{Salama:2015:CFP:2723372.2749437}
{\sc Salama, A., Binnig, C., Kraska, T., and Zamanian, E.}
\newblock Cost-based fault-tolerance for parallel data processing.
\newblock In {\em Proceedings of the 2015 ACM SIGMOD International Conference
  on Management of Data\/} (New York, NY, USA, 2015), SIGMOD '15, ACM,
  pp.~285--297.

\bibitem{flint.Sharma:2016:FBD:2901318.2901319}
{\sc Sharma, P., Guo, T., He, X., Irwin, D., and Shenoy, P.}
\newblock Flint: Batch-interactive data-intensive processing on transient
  servers.
\newblock In {\em Proceedings of the Eleventh European Conference on Computer
  Systems\/} (New York, NY, USA, 2016), EuroSys '16, ACM, pp.~6:1--6:15.

\bibitem{2017sigmetrics:exosphere}
{\sc Sharma, P., Irwin, D., and Shenoy, P.}
\newblock Portfolio-driven resource management for transient cloud servers.
\newblock {\em Proc. ACM Meas. Anal. Comput. Syst. 1}, 1 (June 2017),
  5:1--5:23.

\bibitem{SpotCheck.Sharma:2015:SDD:2741948.2741953}
{\sc Sharma, P., Lee, S., Guo, T., Irwin, D., and Shenoy, P.}
\newblock Spotcheck: Designing a derivative iaas cloud on the spot market.
\newblock In {\em Proceedings of the Tenth European Conference on Computer
  Systems\/} (New York, NY, USA, 2015), EuroSys '15, ACM, pp.~16:1--16:15.

\bibitem{2015socc:spoton}
{\sc Subramanya, S., Guo, T., Sharma, P., Irwin, D., and Shenoy, P.}
\newblock Spoton: A batch computing service for the spot market.
\newblock In {\em Proceedings of the Sixth ACM Symposium on Cloud Computing\/}
  (New York, NY, USA, 2015), SoCC '15, ACM, pp.~329--341.

\bibitem{YARN.Vavilapalli:2013:AHY:2523616.2523633}
{\sc Vavilapalli, V.~K., Murthy, A.~C., Douglas, C., Agarwal, S., Konar, M.,
  Evans, R., Graves, T., Lowe, J., Shah, H., Seth, S., Saha, B., Curino, C.,
  O'Malley, O., Radia, S., Reed, B., and Baldeschwieler, E.}
\newblock Apache hadoop yarn: Yet another resource negotiator.
\newblock In {\em Proceedings of the 4th Annual Symposium on Cloud Computing\/}
  (New York, NY, USA, 2013), SOCC '13, ACM, pp.~5:1--5:16.

\bibitem{Yan:2016:TTC:2987550.2987576}
{\sc Yan, Y., Gao, Y., Chen, Y., Guo, Z., Chen, B., and Moscibroda, T.}
\newblock Tr-spark: Transient computing for big data analytics.
\newblock In {\em Proceedings of the Seventh ACM Symposium on Cloud
  Computing\/} (New York, NY, USA, 2016), SoCC '16, ACM, pp.~484--496.

\bibitem{2017eurosys:pado}
{\sc Yang, Y., Kim, G.-W., Song, W.~W., Lee, Y., Chung, A., Qian, Z., Cho, B.,
  and Chun, B.-G.}
\newblock Pado: A data processing engine for harnessing transient resources in
  datacenters.
\newblock In {\em Proceedings of the Twelfth European Conference on Computer
  Systems\/} (New York, NY, USA, 2017), EuroSys '17, ACM, pp.~575--588.

\bibitem{2010cloud:spotcheckpoint}
{\sc {Yi}, S., {Kondo}, D., and {Andrzejak}, A.}
\newblock Reducing costs of spot instances via checkpointing in the amazon
  elastic compute cloud.
\newblock In {\em 2010 IEEE 3rd International Conference on Cloud Computing\/}
  (July 2010), pp.~236--243.

\bibitem{Zaharia:2010:DSS:1755913.1755940}
{\sc Zaharia, M., Borthakur, D., Sen~Sarma, J., Elmeleegy, K., Shenker, S., and
  Stoica, I.}
\newblock Delay scheduling: A simple technique for achieving locality and
  fairness in cluster scheduling.
\newblock In {\em Proceedings of the 5th European Conference on Computer
  Systems\/} (New York, NY, USA, 2010), EuroSys '10, ACM, pp.~265--278.

\bibitem{Zaharia:2010:SCC:1863103.1863113}
{\sc Zaharia, M., Chowdhury, M., Franklin, M.~J., Shenker, S., and Stoica, I.}
\newblock Spark: Cluster computing with working sets.
\newblock In {\em Proceedings of the 2Nd USENIX Conference on Hot Topics in
  Cloud Computing\/} (Berkeley, CA, USA, 2010), HotCloud'10, USENIX
  Association, pp.~10--10.

\end{thebibliography}

\theendnotes

\end{document}